\documentclass{article}
\usepackage{latexsym}
\usepackage{amsmath}
\usepackage{amssymb}
\usepackage{url}

\newtheorem{exmp}{Example}
\newcommand{\DC}{{\tt{DC}}}
\newcommand{\LC}{{\tt{RLC}}}
\newcommand{\Co}{{\tt{C}}}
\newcommand{\dec}{{\tt{dec}}}
\newcommand{\inc}{{\tt{inc}}}
\newcommand{\pgarltpga}{{\tt{pgarl2pgau}}}
\newcommand{\pgarltpgap}{{\tt{pgarl2pga}}}
\newcommand{\auta}{{\tt{pgau2pga}}}
\newcommand{\lxta}{{\tt{pglx2pga}}}
\newcommand{\rlc}{{\tt{rlc}}}
\newcommand{\use}[1]{\slash_{\!#1}\:}
\newcommand{\rlh}{{\tt{x}\{}}
\newcommand{\rlt}{{\}\tt{x}}}
\newcommand{\setc}{{\tt{set}}}

\newcommand{\uni}{{\tt{u}}}

\newcommand{\st}{\mathsf{S}}
\newcommand{\di}{\mathsf{D}}
\newcommand{\ser}{{\mathcal S}}

\newcommand{\nat}{\Nat}
\newcommand{\Nat}{{\mathbb N}}
\newcommand{\tr}{\true}
\newcommand{\true}{\mathsf{true}}
\newcommand{\fa}{\false}
\newcommand{\false}{\mathsf{false}}
\newcommand{\less}{\sqsubseteq}

 \newdimen\boxwdplusemdimen
  \def\arrow#1{{
    \boxwdplusemdimen=1em%
    \setbox0=\hbox{$\scriptstyle#1$}%
    \advance\boxwdplusemdimen by \wd0\relax%
    \ifdim\boxwdplusemdimen<16.11119pt%
      \boxwdplusemdimen=16.11119pt%
    \fi%
    \buildrel{#1}\over%
      {\setbox1=\hbox to \boxwdplusemdimen{\rightarrowfill}%
    \ht1=0.35em\relax\box1}% %Was .3em
  }}
  \newcommand{\step}[1]{{\ensuremath{\mathbin{\arrow{#1}}}}}

\begin{document}
\title{Projection semantics for rigid loops
\thanks{An earlier version
of this paper appeared as report PRG0604, Section 
Software Engineering, Informatics Institute,
Faculty of Science, University of Amsterdam.}
}
\author{Jan A. Bergstra$^{1,2}$ \and Alban Ponse$^{1}$
\\[1ex]
\small
\begin{tabular}{l}
${}^1$University of Amsterdam, Section Software Engineering,
\\
Kruislaan 403, 1098~SJ~Amsterdam, The Netherlands\\[1ex]
${}^2$Utrecht University, Department of Philosophy,
\\
Heidelberglaan 8, 3584~CS~Utrecht, The Netherlands
\end{tabular}
}

\date{6 July 2007}
\maketitle
\begin{abstract}
A rigid loop is a for-loop with a counter not accessible to the loop body or
any other part of a program. Special instructions for rigid loops are introduced
on top of the syntax of the program algebra PGA. Two different semantic 
projections are provided and proven equivalent. One of these is taken to have 
definitional status on the basis of two criteria: `normative semantic adequacy' 
and `indicative algorithmic adequacy'.

\end{abstract}

\section{Introduction}
In this paper we extend the program algebra PGA \cite{BL02} 
with several new instructions to deal with so-called rigid loops.
Rigid loops are just program fragments that impose the repetition of a 
body a fixed number of times.
%% (in the `normal' case at least).
Rigid loops are very limited in expressive power. Indeed
finite state PGA-programs with rigid loops can be projected into 
equivalent finite state PGA-programs without rigid loops at the 
cost of a combinatorial explosion in length. Like non-recursive 
procedures, rigid loops may be of use when investigating options 
for compiler writing for specific processor architectures. Our specific 
motivation to consider rigid loops arose when studying the potential
gains that may arise from microthread multiplexing
on a single pipelined instruction processing architecture. Following
Jesshope \emph{et al.} in \cite{BJ,J}
loops and nested loops can be usefully split
into microthreads which then may be scheduled either in a multiplexed
fashion on a single pipeline, in an attempt to make use of the unavoidable 
clock cycles 
in which a single thread on the pipeline features stalling, or concurrently
on parallel pipelines on a multiple pipeline architecture in order to
maximize the processing speed for an originally sequential program. 
Rigid loops have the advantage of simplifying dependency analysis,
thus shedding more easily light on what one might hope to achieve. As
it turns out rigid loops are quite interesting even without applications 
like the one just mentioned in mind as a case study for projection semantics. 
Projection semantics has been advocated in \cite{BL02} as a formal 
modeling technique close to programmers intuitions. Rigid loops can be 
easily provided with a projection semantics at the cost of a combinatorial
explosion in program length. Here it will be argued that this is not the most
appropriate way to deal with this issue and another style of projection
which avoids this drastic blow-up in program length is provided
and proven semantically equivalent but algorithmically more natural. 

Although \cite{BL02} provides a clear statement on the objectives and
merits of projection semantics, it fails to provide a methodology which
scales to full size program notations by its exclusive focus on semantic 
issues. Projection semantics provides the meaning of a program notation,
say PGLX, by means of a mapping $\lxta$ from PGLX to PGA which 
assigns to each entity in PGLX a program object (i.e.,
an element of a program algebra, in this case PGA). The program objects
used are finite or infinite instruction streams, over a limited set
of primitive
instructions which goes with the program algebra. As a semantic strategy
projection semantics is independent of this particular program algebra,
but we will use PGA because it works and it allows for a very slow build
up of features, thus permitting a very gradual growth in expressiveness.
The key dogma of projection semantics is that an entity is a program by
either being or representing an instruction stream. Instruction streams
are program objects, i.e., mathematical entities that stand for programs.
Thus a projection explains how some entity can be considered an 
instruction stream and only by explaining (by way of a projection) what
instruction stream an entity represents it can be considered a program.
It is more precise always to speak of a program representation rather than
of a program but because that is very uncommon the term `program'
is used also in cases that a projection does not speak for itself.

Until a projection has been fixed for an entity it is is a candidate program
rather than a program. Only by fixing its projection into an instruction stream
a candidate program becomes a program, comparable to how 
a document becomes legally binding by the addition of relevant 
signatures, locations and dates. We do not accept the conventional viewpoint
that a program can be given a new meaning. Rather a candidate program can
be made to stand for another program by changing its projection, just
as a contract changes when one modifies the signatures. Candidate 
programs may have a quite convincing syntax suggesting meaning without
further ado. We believe that this is \emph{never} actually true. The operational
meaning of candidate programs always requires detailed description
covering a variety of circumstances. Now `projection semantics' as a style of
providing programming language semantics will have to deal with many
notations that are already in
% practical
use and that may have the status of
candidate programs from the perspective of program algebra based
projection semantics, but for which quite satisfactory semantic descriptions
have been found by means of other techniques. Here we are dealing with
providing projection semantics for `known' program notations and the
question may arise as to which semantic description technique is 
most effective. 

Claiming definitional status for a projection for a program
notation that has been given a semantic description already is clearly
problematic.
Therefore the claim can go no further than that a projection might be 
considered to have normative strength semantically, under the hypothesis
that it would be the only description at hand, accepting that in many
cases it will have not have definitional status simply
because other definitions have 
that status already. Such a projection, for a known and well specified program
notation will be called a reconstruction projection semantics in order to 
acknowledge that a definitional status is not claimed. This leads to the 
position that for Pascal one may achieve no more than a reconstruction 
projection semantics while for Perl a projection semantics might still be achievable.

For new or unknown notations, however, whether useful or not, 
a projection 
can be claimed to contain primary semantic information which by definition
cannot be validated or verified against any other description, because of 
its normative nature. Of course validation is possible: by means of a projection 
semantics an operational meaning is assigned to syntactic constructs 
(assuming a string based source language) in a candidate program notation. Because
the syntax of this candidate program notation is itself a matter of meticulous
design the operational meaning should make best  possible use of the syntax that
has been made available. If a projection prescribes an unintelligible meaning to
a construct that might have been given a clear and useful meaning instead
a design error has occurred which can and probably should be repaired. 

Returning to the issue that known program notations cannot be given a 
projection semantics the following solution to this somewhat philosophical 
issue can be found. For projection
semantics as a topic of investigation this philosophical matter is simply 
solved by always using slightly unconventional syntax (however marginal 
the differences) such that the setting establishes
a new syntax which is given a meaning
for the first and therefore definitive time. The ability of a projection for a candidate 
program notation to serve as a carrier of intended semantic information is
termed \emph{normative semantic adequacy}. Normative semantic adequacy
does not come for free: it requires that comprehensible projections into
comprehensible programs are used to provide a realistic, suggestive and
useful meaning (in terms of instruction streams) for new syntax. Usually
a projection will be into a program notation that has been provided
with a projection semantics already thus giving rise to chains of projections.

Besides normative semantic adequacy one also expects a projection 
to represent an indication (or model) of how the actual processing of
a (candidate) program might in practice proceed. Exponential or even 
polynomial blow-up of the size of an entity during its projecting
transformation are signs that \emph{indicative algorithmic adequacy}
has not been achieved.

A projection for a programming notation feature which enjoys
both normative semantic adequacy and indicative algorithmic adequacy 
is called a defining projection. If it uses some services of type T it will be
called a T service based defining projection. Using this terminology
 we will develop in this paper a rigid
loop counter service based defining projection for PGArl 
(PGA with rigid loops).

The further content of this paper is divided into four parts:
in Section~\ref{Threads} we formally introduce threads and
services.
In Section~\ref{Bas} we introduce the program algebra PGA, thread
extraction and a further extension of PGA.
%% ,  and its string syntax PGLA, its extension PGLB and for each of these
In Section~\ref{PGArl} we extend PGA 
%% , PGLArl and PGLBrj
with rigid loops to PGArl,
including two forms of projection semantics. It is clarified
that the projection semantics making use of decreasing loop 
counters enjoys both normative semantic adequacy and indicative 
algorithmic adequacy and that the pure projection into PGA fails
for the second criterion.
% Then, in Section~\ref{Eas} we introduce execution architectures 
% based on Maurer computers. It is shown that in this case a language 
% with rigid loops only and without repetition has universal expressive power.
The paper is ended with some conclusions in Section~\ref{Conc}.

\section{Threads and Services}\label{Threads}
The behavior of programs under execution is modelled by
\emph{threads}.
In this section we introduce thread algebra.
Then we introduce services, devices that can be
\emph{used} by a thread in order to increase expressiveness.

\subsection{Thread algebra}
{Basic thread algebra}, or BTA for short,
is intended for the description of sequential program behavior
(see \cite{BM05}; in \cite{BL02} BTA is
introduced as \emph{basic polarized process algebra}).
Based on a finite set $A$ of \emph{actions} 
it has the following constants and operators:
\begin{itemize}
\item the \emph{termination} constant $\st$,
\item the \emph{deadlock} or \emph{inaction} constant $\di$,
\item for each $a\in A$, a binary \emph{postconditional composition}
operator $\_  \unlhd a \unrhd \_$.
\end{itemize}
We use \emph{action prefixing} $a \circ P$ as an abbreviation for
$P \unlhd a \unrhd P$ and take $\circ$ to bind strongest.
Furthermore, for $n\in \nat$ we define $a^n\circ P$ by $a^0\circ P=P$ and
$a^{n+1}\circ P=a\circ (a^n\circ P)$.

The operational intuition behind thread algebra is that each  action
represents a request to be processed by the execution
environment.
% of a thread which may involve a change of state of this environment.
At completion of the processing of the request, the environment
produces a reply value $\tr$ or $\fa$
to the thread under execution and may undergo a change of state.
The thread $P \unlhd a \unrhd Q$ will
then proceed as $P$ if the processing of $a$ yielded the reply $\tr$
indicating successful processing, and it will proceed as $Q$ if
the processing of $a$ yielded the reply $\fa$.

BTA can be equipped with a partial order and an
\emph{approximation operator}.
\begin{enumerate}
\item $\less $ is the
partial ordering on BTA generated by the clauses
\begin{enumerate}
\item for all $P\in \textrm{BTA}$, $\di\less P$, and
\item for all $P_1,P_2,  Q_1, Q_2\in \textrm{BTA}$, $a \in A$,
\[P_1\less Q_1\ \&\ P_2 \less Q_2 \Rightarrow P_1 \unlhd a
\unrhd P_2 \less Q_1 \unlhd a \unrhd Q_2.\]
\end{enumerate}
\item
$\pi: \nat \times\textrm{BTA} \rightarrow \textrm{BTA}$
is the approximation operator
determined by the equations
\begin{enumerate}
\item for all $P \in \textrm{BTA}$, $\pi(0,P) = \di$,
\item for all $n\in \nat$, $\pi(n+1,\st) = \st,\,\, \pi(n+1,\di) = \di$, and
\item for all $P,Q \in \textrm{BTA}, n\in \nat$, \[\pi(n+1,P\unlhd a \unrhd Q) =
\pi(n,P)\unlhd a \unrhd \pi(n,Q).\]
\end{enumerate}
We further write $\pi_n(P)$ instead of $\pi(n,P)$.
\end{enumerate}
The operator $\pi$ finitely approximates every thread in
BTA. That is, for all $P\in \textrm{BTA}$,
\[ \exists n\in \nat \ \pi_0(P)\less \pi_1(P)\less \cdots \less \pi_n(P)
=\pi_{n+1}(P) = \cdots = P.\]
%The value $n$ is in this case called the \emph{depth} of $P$.
Threads can be finite or infinite.
Following the metric theory  of \cite{BZ82}
as the basis of processes in \cite{BK84}, BTA
has a completion BTA$^\infty$ which comprises also infinite threads.
Standard properties of the completion technique yield that we may take
BTA$^\infty$ as the cpo consisting of all so-called
\emph{projective} sequences.  That is,
\[
\textrm{BTA}^\infty = \{(P_n)_{n\in \nat} \mid
\forall n \in \nat\  (P_n\in  \textrm{BTA}\ \&\ \pi_n(P_{n+1})=P_n) \}
\]
with
\[
(P_n)_{n\in \nat} \less (Q_n)_{n\in \nat} \Leftrightarrow
\forall n \in \nat \ P_n \less Q_n
\]
and
\[
(P_n)_{n\in \nat} = (Q_n)_{n\in \nat} \Leftrightarrow
\forall n \in \nat \ P_n = Q_n.
\]
(For a detailed account of this construction see \cite{BB03}.)

Let $I=\{1,...,n\}$ for some $n>0$.
A \emph{finite linear recursive specification}
over BTA is a set of equations
\[X_i=t_i(\overline X)\]
for $i\in I$ with $\overline X=X_1,...,X_n$ and all $t_i(\overline X)$
of the form
$\st$, $\di$, or $X_{i_l}\unlhd a_i\unrhd X_{i_r}$ for $i_l,i_r\in I$ and
$a_i\in A$.
In $\textrm{BTA}^\infty$,
finite linear recursive specifications represent
continuous operators having as unique fixed points
\emph{regular} threads,
i.e., threads which can only reach finitely many states.

\begin{exmp}\label{exproc}
Let $n>0$.
The regular thread
$a^n\circ \di$ is the fixed point for $X_1$ in the specification
$$ \{X_i=a\circ X_{i+1}\mid i=1,...,n\}\cup\{X_{n+1}=\di\}. $$
The regular thread
$a^n\circ \st$ is the fixed point for $X_1$ in
$$ \{X_i=a\circ X_{i+1}\mid i=1,...,n\}\cup\{X_{n+1}=\st\}. $$
Both these threads are finite.

The infinite regular thread $a^\infty$
is the fixed point for $X_1$ in the specification
$\{X=a\circ X\}$ and corresponds
to the projective sequence $(P_n)_{n \in \nat}$ with $P_0 =\di$ and
$P_{n+1}=a\circ P_n$.

Observe that e.g.\ $a^n \circ \di \less a^n \circ \st$,
$a^n \circ \di \less a^\infty$ but $a^n \circ \st \not\less a^\infty$.
\end{exmp}

For the sake of simplicity, we shall often define regular
threads by providing only one or more equations.
For example, we say that $P=a\circ P$ defines a regular
thread with name $P$ (so $P=a^\infty$ in this case).

We end this section with the observation that
for regular threads $P$ and $Q$, $P\less Q$ is decidable.
Because one can always take the disjoint
union of two recursive specifications,
it suffices to argue that $P_i\less P_j$ in
\[
P_1=t_1(\overline P),...,P_n=t_n(\overline P)
\]
is decidable.
This follows from the assertion
\begin{equation}\label{imp}
\forall i,j\leq n\ \pi_n(P_i)\less\pi_n(P_j)\Leftrightarrow P_i\less
P_j,
\end{equation}
where $\pi_l(P_k)$ is defined by $\pi_l(t_k(\overline P))$,
because $\less$ is decidable for finite threads.
Without loss of generality, assume $n>1$.
To prove \eqref{imp}, observe that $\Leftarrow$ follows by
definition of regular threads. For the reverse,
choose $i,j$ and assume that $\pi_n(P_i)\less\pi_n(P_j)$.
Suppose $P_i\not\less P_j$,
then for some $k>n$, $\pi_k(P_i)\not\less\pi_k(P_j)$ while
$\pi_{k-1}(P_i)\less\pi_{k-1}(P_j)$.
So there exists a trace of length $k$ from $P_i$ of the form
\[P_i\step{a_{\tr}}P_{i'}\step{b_{\fa}}...\]
that is not a trace of $P_j$, while by the assumption
the first $n$ actions are a trace of $P_j$.
These $n$ actions are connected by $n+1$ states, and since
there are only $n$ different states $P_1, ...,P_n$, a repetition
occurs in this sequence of states.
So the trace witnessing $\pi_k(P_i)\not\less \pi_k(P_j)$
can be made shorter,
contradicting $k$'s minimality and hence the supposition.
Thus $P_i\less P_j$.
Consequently, also $P=Q$ (i.e., $P\less Q$ and $Q\less P$)
is decidable for regular threads $P$ and $Q$.

\subsection{Services}\label{SM}
A \emph{service} is a pair \(\langle\Sigma, F\rangle\)
consisting of a set $\Sigma$ of
so-called \emph{co-actions} and a \emph{reply function} $F$.
This reply function is
a mapping that gives for each finite sequence of co-actions
from $\Sigma$ a reply value $\tr$ or $\fa$.
Services were introduced in~\cite{BP02} under the name ``state
machines''.

\begin{exmp}\label{counter}
A \emph{down counter} or \emph{loop counter}
is a service $\DC=\langle\Sigma, F\rangle$ with
$\Sigma=\{\dec,\setc{:}n\mid n\in I\}$ consisting of the 
decrease and set co-actions for some $I\subseteq\nat$ and
the reply function $F$ which replies $\tr$ to $\setc{:}n$ while
setting $\DC$ to value $n$, and $\tr$ to $\dec$ if
$\DC$'s value is positive while decreasing its current
value, and $\fa$ to $ \dec$ if and only if the counter is zero.
The initial value of $\DC$ is zero and usually $I$ will be an
initial segment of $\nat$.
\end{exmp}
Down counters (also known as timer units) are crucial components of most
embedded systems and included in many microcontrollers (see e.g.
\cite{BB02}).
Below, we return to this example.

In order to provide a specific description of the interaction
between a thread and a service, we will use for actions the general notation
${c}.{a}$ where $c$ is the so-called
\emph{channel} or \emph{focus}, and $a$ is the
{co-action}.
For example, 
$c.\inc$ is the action which increases a counter via channel
$c$.
This interaction is is defined with help of the \emph{use
operator} $\use{}$.
For a service $\ser=\langle\Sigma, F\rangle$,
a finite thread $P$ and a channel $c$, the
defining rules for $P\use c \ser$ (the thread \emph{$P$ using the service
$\ser$ via channel $c$}) are:
\[\begin{array}{rcl}
\st\use c \ser &=& \st,\\
\di\use c \ser &=& \di, \\
(P\unlhd c'.a\unrhd Q)\use c\ser &=&
(P\use c \ser)\unlhd c'.a \unrhd (Q\use c \ser)
\text{ if } \mathtt c'\neq \mathtt c,\\
(P\unlhd c.a\unrhd Q)\use c \ser &=&
P\use c  \ser'\text{
if $a\in\Sigma$ and } F(a)=\tr,\\
(P\unlhd c.a\unrhd Q)\use c \ser &=&
Q\use c  \ser'\text{
if $a\in\Sigma$ and } F(a)=\fa,\\
(P\unlhd c.a\unrhd Q)\use c \ser &=& \di\text{
if $a\not\in\Sigma$.}
\end{array}\]
where $\ser'= \langle\Sigma, F'\rangle$ with $F'(\sigma)
=F(a\sigma)$ for all co-action sequences $\sigma\in \Sigma^+$.
The use operator is expanded to infinite threads $P$ by defining
\[
P\use c \ser = \bigsqcup_{n\in \nat}\pi_n(P)\use c \ser.
\]
(Cf.\ \cite{BB}.)
As a consequence, $P\use c \ser=\di$ if for any $n$, $\pi_n(P)\use c \ser=\di$.
Of course, repeated
applications of the use operator bind to the left, thus
\[P\use{c0} \ser_0\use{c1}\ser_1=(P\use{c0} \ser_0)\use{c1}\ser_1.\]

We end this section with an example on the use of a service,
showing that non-regular threads can be specified with
infinite state services.

\begin{exmp}\label{inclu}
We may extend the down counter defined in Example~\ref{counter}
to a \emph{full counter} $\Co$ by including co-actions
$\inc$ (increase)
which always yield reply $\tr$ while increasing the counter
value.
Now let $\{a,b,c.\inc,c.\dec\}\subseteq A$.
We write $\Co(n)$ for a counter with value $n\in\Nat$, so $\Co=\Co(0)$.
By the defining equations for the use operator it follows that
for any thread $P$,
\[(c.\inc\circ P)\use c \Co(0)=P\use c \Co(1),
\]
and  $\forall n\in\Nat$,
$({c}.{inc}\circ P)\use c \Co(n)=P\use c \Co(n+1)$.
Furthermore, it easily follows that
\[(P\unlhd{c}.\dec\unrhd \st)\use c \Co(n)=
\begin{cases}\st & \text{ if } n=0,\\
P\use c \Co(n-1) & \text{ otherwise.}
\end{cases}
\]
Now consider the regular thread $Q$
defined by\footnote{Note that a \emph{linear}
recursive specification of $Q$ requires
(at least) five equations.}
\begin{eqnarray*}
Q&=&({c}.\inc\circ Q)\unlhd a\unrhd R,\\
R&=&b\circ R\unlhd{c}.{\dec}\unrhd \st.
\end{eqnarray*}
Then 
\begin{eqnarray*}
Q\use c \Co(0)&=&(({c}.{\inc}\circ Q)\unlhd a\unrhd R)\use c \Co(0)\\
&=&(Q\use c \Co(1))\unlhd a\unrhd (R\use c \Co(0),
\end{eqnarray*}
and for all $n\in\Nat$,
$Q\use c \Co(n)= (Q\use c \Co(n+1))\unlhd a\unrhd (R\use c \Co(n)$.
It is not hard to see that $Q\use c \Co(0)$ 
is an infinite thread with the property that for all $n$,
a trace of $n+1$ $a$-actions produced by $n$ positive
and one negative reply on $a$ is followed by $b^n\circ\st$.
This yields an \emph{irregular} thread:
if $Q\use c \Co(0)$
were regular, it would be a fixed point of some finite linear recursive
specification, say with $k$ equations.
But specifying a trace $b^k\circ\st$ already requires $k+1$ linear
equations $X_{1}=b\circ X_{2},...,
X_{k}=b\circ X_{k+1},X_{k+1}=\st$, which contradicts the
assumption. So $Q\use c \Co(0)$ is not regular.
\end{exmp}

\section{Programs and Program Algebra}
\label{Bas}
In this section we introduce the program algebra PGA
(see \cite{BL02}) and discuss its relation with thread algebra. 
Furthermore, we shortly discuss the unit instruction operator.

\subsection{PGA, basics of program algebra}\label{basics}
Given a thread algebra with actions in $A$, we now consider
the actions as so-called \emph{basic instructions}.
The syntax of PGA
% (PGA$(A)$ is a more precise notation)
has the following primitive instructions as constants:
\begin{description}
\item[\it Basic instruction] $a\in A$.
It is assumed that upon the execution of a basic instruction, the
(executing) environment provides an answer $\true$ or $\false$. However, in the
case of a basic instruction, this answer is not used for program control.
 After execution of a basic instruction, the next instruction (if any)
will be executed; if there is no next instruction, inaction will occur.
\item[\it Positive/negative test instruction] $\pm a$ for $a\in A$.
A positive test instruction $+a$ executes 
like the basic instruction $a$. Upon $\false$, the program skips its next
instruction and continues with the instruction thereafter; upon $\true$
the program executes its next instruction. For a negative test instruction
$-a$, this is reversed: upon $\true$, the program skips its next instruction
and continues with the instruction thereafter; upon $\false$ the program
executes its next instruction.
If there is no subsequent instruction to be executed, inaction occurs.
\item[\it Termination instruction] $!$. This instruction prescribes successful
termination.
\item[\it Jump instruction] $\#k$ ($k\in\Nat$). This instruction prescribes
execution of the program to jump $k$ instructions forward; if there is no
such instruction, inaction occurs. 
In the special case that $k=0$, this prescribes a jump to the instruction
itself and inaction occurs, in the case that $k=1$ this jump acts as a
\emph{skip} and the next instruction is executed. In the case that the
prescribed instruction is not available, inaction occurs.
\end{description}

PGA-terms are composed by means of \emph{concatenation}, notation
$\_;\_$, and
\emph{repetition}, notation $(\_)^\omega$. 
Instruction sequence congruence for
PGA-terms is axiomatized by the axioms PGA1-4 in Table~\ref{Tab1}.
Here PGA2 is an axiom-\emph{scheme}: for each $n>0$,
$(X^n)^\omega=X^\omega$, where $X^1=X$ and $X^{k+1}=X;X^k$.
A closed PGA-term is often called a PGA-program.

\begin{table}[htbp]
\caption{Axioms for PGA's instruction sequence congruence}
\centering
$
~\begin{array}{rcll}
\hline \\[-2mm]
\hspace{1.3cm}(X;Y);Z &=& X;(Y;Z)&\mathrm{(PGA1)}\hspace{1.2cm}\\
(X^{n})^{\omega} &=& X^{\omega}\quad\text{for}\quad n>0\quad\quad&
\mathrm{(PGA2)}\\
X^{\omega};Y &=& X^{\omega}&
\mathrm{(PGA3)}\\
(X;Y)^{\omega} &=& X;(Y;X)^{\omega}&
\mathrm{(PGA4)}\\[2mm]
\hline
\end{array}
$
\label{Tab1}
\end{table}

From the axioms PGA1-4 one easily derives \emph{unfolding}, i.e.,
\[X^\omega=X;X^\omega.\]
Furthermore, 
each PGA-program can be rewritten into an instruction equivalent
\emph{canonical form}, i.e., a closed term of the form $X$
or $X;Y^\omega$ with $X$ and $Y$ not containing repetition.
This also follows from the axioms in Table~\ref{Tab1}.

We will often use basic instructions in so-called
\emph{focus.method} notation, i.e., basic instructions of the form
\[f.m\]
where $f$ is a focus (channel name) and $m$ a method name.
The $m$ here is sometimes called a
\emph{service-instruction} because it
refers to the use of some service, and is 
related with a co-action as defined in Section~\ref{SM}.
Two examples of instructions in focus.method notation are
${c}.{\inc}$ and ${c}.{\dec}$,
related with the actions controlling a counter discussed
in Example~\ref{inclu}.
In the next section we will relate all
basic and test instructions to
the actions of a thread; this is called \emph{thread
extraction}.

\subsection{Thread extraction: from PGA to thread algebra}
The \emph{thread extraction} operator $|X|$ assigns a 
thread to program object $X$.
Thread extraction is defined by the thirteen equations in
Table~\ref{Tab3},
where $a\in A$ and $u$ is a primitive instruction.

\begin{table}[htbp]
\caption{Equations for thread extraction on PGA}
\centering
$
~\begin{array}{rcl} 
\hline
\\[-2mm]
|!| &=& \st\\
|a| &=& a\circ \di\\
|{+a}| &=& a \circ \di\\
|{-a}| &=& a \circ \di\\[2mm]
|!;X| &=& \st\\ 
|a;X| &=& a\circ |X|\\
|{+a};X| &=& |X|\unlhd a \unrhd |\#2;X| \\
|{-a};X| &=& |\#2;X|\unlhd a \unrhd |X|\\[2mm]
|\#k|&=&\di\\
|\#0;X| &=& \di\\
|\#1;X| &=& |X|\\
|\#k+2;u| &=& \di\\
\hspace{15mm}|\#k+2;u;X| &=& |\#k+1;X|
\hspace{15mm}\\[2mm]
\hline
\end{array}$
\label{Tab3}
\end{table}

Some examples: 
\begin{eqnarray*}
|(\#0)^{\omega}|&=&|\#0;(\#0)^{\omega}|
%\\&=&
=\di, \\[2mm]
|{-a};b;c|
&=&|\#2;b;c|\unlhd a \unrhd |b;c|\\
&=&|\#1;c|\unlhd a \unrhd b\circ |c|\\
&=&|c|\unlhd a \unrhd b\circ c\circ \di\\
&=&c\circ \di\unlhd a \unrhd b\circ c\circ \di.
\end{eqnarray*}
In some cases, these equations can be applied from left to right
without ever generating any behavior, e.g.,
\[\begin{array}{l}
|(\#2;a)^{\omega}|=|\#2;a;(\#2;a)^{\omega}|=|\#1;(\#2;a)^{\omega}|
=|(\#2;a)^{\omega}|=\ldots
\end{array}\]
In such cases, the extracted thread is defined as $\di$.

It is also possible that thread extraction yields an infinite recursion,
e.g., \[|a^\omega|=|a;a^\omega|=a\circ|a^\omega|\]
(in the previous section we denoted this thread by
$a^\infty$). If the behavior of $X$ is infinite, it
is regular and can be represented by a (linear) recursive
specification, e.g.,
\begin{eqnarray*}
|(a;+b;\#3;{-}b;\#4)^\omega|=P \text{ in }
P&=&a \circ (P \unlhd b\unrhd Q),\\
Q&=&P \unlhd b\unrhd Q.
\end{eqnarray*}
It follows easily that any PGA-program defines a regular thread, 
and conversely, each regular thread can be defined in PGA:
linear equations of the form $X=\st$ or $X=\di$ can be defined by
instructions $!$ and $\#0$, respectively, and
a linear equation
\[X=Y\unlhd a\unrhd Z\]
can be associated with a triple
\({+}a;\#k;\#l.\)
Connecting these program fragments in a repetition and
instantiating the jump counters $k$ and $l$ with the appropriate
values then yields a PGA-program that defines a solution for the
first equation. A typical example:
\[
\begin{array}{lll}
P_1&=&P_2\unlhd a\unrhd P_2,\hspace{1cm}\\
P_2&=&P_3\unlhd b\unrhd P_1,\\
P_3&=&\di.
\end{array} 
\mapsto\hspace{1cm}
\begin{array}{l}
(+a;\#2;\#1;\\
+b;\#2;\#2;\\
\#0)^\omega.
\end{array}
\]
For PGA-programs $X$ and $Y$ we write
\[X=_{be} Y\]
if $X$ and $Y$ are behaviorally equivalent (i.e., have
the same behavior). Behavior equivalence is not a congruence,
e.g., $\#0=_{be}\#1$ but $\#0;a\neq_{be}\#1;a$.
Finally, for a PGA-program $X$ we define 
\[X\use c \ser\]
as the program with behavior $|X|\use c \ser$, thus $|X\use c
\ser|=|X|\use c \ser$.

\subsection{PGAu, PGA with unit instruction}
In \cite{BL02} the \emph{unit instruction operator}, notation $\uni(\_)$
is introduced.
This operator wraps a program fragment into
a single unit: if X is a program, then $\uni(X)$ is a unit that upon
execution behaves as $X$, but that counts as a single 
instruction in any context.
A typical example is 
\[+a;\uni(b^\omega);c\]
which behaves as 
\[|b^\omega|\unlhd a \unrhd |c|.\]
A PGA-program that defines the same thread as the above example is for
instance
\[+a;(\#2;\#3;b;\#3;c;\#0)^\omega.\]
Typically, a jump to a non-starting position in a unit is not
possible, while a jump out of a unit can occur in any
position of its body. As an example,
\[+a;\#3;\uni(+b;\#3;c);d;e\]
defines the same thread as $+a;\#5;+b;\#3;c;d;e$,
i.e.,
\[|e|\unlhd a \unrhd(|e|\unlhd b\unrhd |c;d;e|).\]
Incorporating the unit instruction operator in PGA, notation PGAu,
does not increase the expressive power.
In this paper we shall make a modest use of the unit instruction
operator and we refrain from describing the projection semantics
for PGAu as defined in \cite{Pon}.
\footnote{This formal semantics is implemented in the PGA
Toolset \cite{Diertens} and | including an application of "jump-optimization"
| yields for the examples above
\begin{eqnarray*}
&&+a;
(\#2;
\#3;
b;
\#5;
c;
\#0)^\omega, \text{ and}
\\
&&+ a;
(\#5;
+ b;
\#3;
c;
d;
e;
\#0
)^\omega \text{ respectively}.
\end{eqnarray*}
}

The projection semantics for PGAu is
defined by a projection function $\auta$ (in \cite{Pon})
on first canonical PGAu-forms, i.e., 
closed terms of the form $X$
or $X;Y^\omega$ with $X$ and $Y$ not containing repetition.
In the particular case
that a program contains no units, these are first canonical
forms in PGA. Furthermore, the projection $\auta$ yields
in all cases
PGA-programs of the form $(u_1;...;u_k)^\omega$ and has
definitional status. 
Consequently, each PGAu-program | and therefore each PGA-program
| can be expressed in this form.
In the next section we will use this property for PGA extended
with rigid loops.

\section{PGA with rigid loops}\label{PGArl}
In this section we add two types of non-primitive instructions to PGA,
thus obtaining PGA with rigid loops. Then we discuss a projection semantics
that maps programs to PGAu using counters. We postulate that this semantics
has definitional status and argue that this is a reasonable proposal by
discussing a ``pure projection''. Finally,
we consider some degenerate examples.

\subsection{PGArl, PGA with rigid loops}
We add two types of non-primitive instructions to PGA,
thus obtaining PGArl, i.e., PGA with rigid loops:
\begin{description}
\item[\it Rigid loop header instruction] $n\rlh$ for
each $n\in\Nat\setminus\{0\}$. 
Examples are $7\rlh$ and $432\rlh$. 
This instruction prescribes an $n$ times repeated
execution of the program fragment until the following complementary rigid
loop closure instruction. During execution of the body, jumps out of it are 
permitted and will end its execution; termination within a loop entails
termination of the whole program and so does a livelock ($\#0$).
A jump into the body of a rigid loop prescribes the execution
of its remaining instructions.

\item[\it Rigid loop closure instruction] $\rlt$. 
This instruction ends the body of a rigid loop.
\end{description}
The idea is that the matching of header and closure instructions is
innermost-outermost: instruction sequences are parsed left-to-right, so a
closure instruction matches the last preceding rigid loop header.

The semantics of PGArl is given by a projection which makes use of an 
intermediate stage involving annotated closure instructions for rigid
loops and annotated jumps out of rigid loops.

\begin{description}
\item[\it Annotated rigid loop closure instruction] $n\rlt m$ for each $n$ and 
$m \in\Nat$. 
This instruction ends the body of a rigid loop with counter value
$n+1$ of which the body has a size of $m$ instructions.
Its execution is best explained in the presence of a separate 
loop counter $\LC$ (cf.\ Example~\ref{counter})
which is  initialised at $n$ before execution 
of the rigid loop and records the number of repetitions still to be done.
Executing the annotated closure instruction then consists of 
$\#1$ if the loop counter $\LC$ has reached value $0$ and otherwise a jump to
the first instruction of the loop body.
These activities must be packed into a single unit in order to
preserve the validity of other jumps elsewhere in the program.

In the case that there is no associated rigid loop header
instruction, the annotation is $0\rlt0$.

\item[\it Annotated jump instruction]
$\#l(j_1,n_1)(j_2,n_2)...(j_k,n_k)$
with $j_i,n_i,k\in\Nat$ for a jump $\#l$ that jumps over $k$
annotated closure instructions
$n_1\rlt m_1,...,$  $n_k\rlt m_k$ at positions $j_1,...,j_k$.
The annotation will be used to reset all concerning loop counters.
\end{description}
%$\rlta$. 

As an example, $3\rlh;a;b;4\rlh;+c;\#4;\rlt;d;\rlt;+e;\#3$ yields the
annotation
\[3\rlh;a;b;4\rlh;+c;\#4(7,3)(9,2);3\rlt 2;d;2\rlt 7;+e;\#3.\]

We start with the case that a PGArl-program is of the form
\[(u_1;...;u_k)^\omega,\]
a form 
which easily facilitates a backward jump to the first
instruction of the body of a rigid loop.
We adopt the following restrictions on $(u_1;...;u_k)^\omega$:
\begin{itemize}
\item each rigid loop header instruction has a complementary
closure instruction,
\item
for each jump instruction $\#m$ it holds that $m<k$ (if not, subtract $k$
sufficiently often),
\item rigid loop closures are not preceded by a test
instruction.
\end{itemize}
For the projection we need first to add the annotations, and then to 
introduce a service for a loop counter attached to each annotated rigid 
loop closure instruction. The closure instruction at position $i$ will 
make use of service $\rlc{:}i$. A loop counter has methods $\setc{:}n$ which 
initialises it to $n$ and $\dec$ which subtracts $1$ if possible while returning 
a reply $\tr$ and otherwise returns the reply $\fa$.

The projected program begins with an initialisation instruction
$\rlc{:}i.\setc{:}c_{i}$ where 
$c_{i}$ is the left annotation of the annotated loop closure
instruction for each rigid loop
that occurs in the candidate program. The loop headers are projected to
$\#1$ and their only role has been to determine the annotations for the closure 
instructions.
Thus, assuming that $u_1;...;u_k$ contains $l$
rigid loops with annotated
closure instructions at positions $i_1,i_2,..., i_l$,
we define
\[\begin{array}{rcl}
\pgarltpga((u_1;...;u_k)^\omega)&=&\rlc{:}i_1.\setc{:}c_{i1};
\rlc{:}i_2.\setc{:}c_{i2};...;\rlc{:}i_l.\setc{:}c_{il};\\
\multicolumn{3}{r}{
(\psi_1(u_1);...;\psi_k(u_k))^\omega
\use {\rlc{:}i_1} \LC_{i_1}
\use {\rlc{:}i_2} \LC_{i_2}
...
\use {\rlc{:}i_l} \LC_{i_l}}
\end{array}\]
with
\[\begin{array}{rcl}
\psi_i(n\rlh)&=&\#1,\\
\psi_i(\#l(j_1,n_1)...(j_m,n_m))&=&\uni(\begin{array}[t]{l}
\rlc{:}j_1.\setc.n_1;\\
...\\
\rlc{:}j_m.\setc.n_m;\#l),\end{array}\\
\psi_{i}(n\rlt m) &=&\uni(\begin{array}[t]{l}
+\rlc{:}i.\dec;\#3;\\
\rlc{:}i.\setc{:}n;\#2;\\
\#k-m),\end{array}
\\
\psi_i(u)&=&u \textrm{ otherwise.}
\end{array}\]
Note that in case $(u_1;...;u_k)^\omega$ does not contain
rigid loop instructions, we have
by definition that
$\pgarltpga((u_1;...;u_k)^\omega)=(u_1;...;u_k)^\omega$.

As a first example,
$(3\rlh;a;b;4\rlh;c;\rlt;d;\rlt;e)^\omega$ yields the
annotated program
\[(3\rlh;a;b;4\rlh;c;3\rlt1;d;2\rlt6;e)^\omega,\]
which yields under $\pgarltpga$ 
%\[\begin{array}{l}
\begin{eqnarray*}
&&\rlc{:}6.\setc{:}2;
\rlc{:}8.\setc{:}3;\\
&&(\#1;a;b;\#1;c;
~\uni(\begin{array}[t]{l}
+\rlc{:}6.\dec;\#3;\\
\rlc{:}6.\setc{:}2;\#2;\\
\#8);
\end{array}
\\
&&~~d;
%\\
~\uni(\begin{array}[t]{l}
+\rlc{:}8.\dec;\#3;\\
\rlc{:}8.\setc{:}3;\#2;\\
\#3);
\end{array}
\\
&&~~e
%\\
)^\omega
\use {\rlc{:}6} \LC_{6}
\use {\rlc{:}8} \LC_{8}
\end{eqnarray*}
%\end{array} \]
and thus defines the thread $P$ given by
$P=(a\circ b\circ c^4\circ d)^3\circ e\circ P$.

As a second example, consider the program
$(a;2\rlh;+b;\#3;\rlt;c;d)^\omega$, thus
\begin{equation*}
(a;2\rlh;+b;\#3(5,1);1\rlt2;c;d)^\omega,
\end{equation*}
which has the
option of ending a rigid loop by jumping out of it:
under
\\
$\pgarltpga$ we obtain
\begin{eqnarray*}
&&\rlc{:}5.\setc{:}1;\\
&&(a;\#1;+b;
~\uni(\rlc{:}5.\setc{:}1;\#3);\\
&&
~~\uni(\begin{array}[t]{l}
+\rlc{:}5.\dec;\#3;\\
\rlc{:}5.\setc{:}1;\#2;\\
\#5);
\end{array}
\\
&&~~c;d
)^\omega
\use {\rlc{:}5} \LC_{5}
\end{eqnarray*}
which defines the thread $P$ given by
\begin{eqnarray*}
P&=&a\circ (d\circ P\unlhd b\unrhd (d\circ P\unlhd
b\unrhd c\circ d\circ P)).
\end{eqnarray*}

For a repetition-free  PGArl-program $u_1;...;u_k$ we define
\[\pgarltpga(u_1;...;u_k)=\pgarltpga(\Phi(u_1;...;u_k)),\]
where the transformation $\Phi$ is given by
\[\begin{array}{l}
\Phi(u_1;...;u_k)= (\phi_1(u_1);...;\phi_k(u_k);\#0;\#0)^\omega,\\
\phi_i(\#n)=\#\min{(n,k+2-i)},\\
\phantom{\#}\phi_i(u)=u\textrm { otherwise.}
\end{array}\]
Here the latter two $\#0$-instructions serve the case that $u_k$ is
a test instruction.

It remains to define
the projection $\pgarltpga$ for first
canonical forms
\[u_1;...;u_k;(v_1;...;v_l)^\omega\]
with $k,l>0$. In this case we may assume that if $u_i=\#m$, then
$m\leq k-i+l$ (otherwise, subtract $l$ sufficiently often).
Similarly, we may assume that if $v_j=\#m$, then $m<l$.
We define
\[\pgarltpga(u_1;...;u_k;(v_1;...;v_m)^\omega)=
\pgarltpga(\Xi(u_1;...;u_k;(v_1;...;v_m)^\omega))\]
with
\[\begin{array}{l}
%\begin{eqnarray*}
\Xi(u_1;...;u_k;(v_1;...;v_m)^\omega)=
(u_1;...;u_k;\xi_1(v_1);...;\xi_m(v_m);\#k;\#k)^\omega,\\
\xi_i(\#n)=\#n+k+2 \text{ if $i+n>m$},\\
\phantom{\#}\xi_i(u)=u \textrm{ otherwise.}
%\end{eqnarray*}
\end{array}\]
This completes the definition of $\pgarltpga$ and we give this projection
\emph{definitional status}. In other words,
the loop counter service based projection $\pgarltpga$
is the \emph{defining projection} for PGArl.

\subsection{Pure projection of rigid loops and definitional
status}
In the previous section we assumed that PGArl-programs satisfy a certain
well-formedness criterion:
\begin{itemize}
\item each rigid loop header instruction has a complementary
closure instruction,
\item
for each jump instruction $\#m$ in $(u_1;...;u_k)^\omega$
it holds that $m<k$ (if not, subtract $k$
sufficiently often),
\item rigid loop closures are not preceded by a test
instruction.
\end{itemize}
Before dealing with programs that are not well-formed, we first discuss
pure projection of well-formed PGArl-programs.

The pure PGA projection $\pgarltpgap$
expands the body of each loop while adapting 
appropriately the jumps that go into the body and that might exit from the body.
Expansion can be defined in a left-to-right order on rigid loop
headers in the following way:
let $X$ be a (possibly empty) sequence of PGA-instructions, 
$u_i$ range over the PGArl-instructions, and let 
$Y$ range over finite (possibly empty)
sequences of PGArl-instructions. Then
\begin{equation}
\label{een}
X;1\rlh;u_1;...;u_k;\rlt;Y=X;\#1;u_1;...;u_k;\#1;Y
\end{equation}
and for all $n>1$,
\begin{equation}
\label{twee}
X;(n+1)\rlh;u_1;...;u_k;\rlt;Y=X';\#1;u_1';...;u_k';\#1;
  n\rlh;u_1;...;u_k;\rlt;Y
\end{equation}
where 
\begin{eqnarray*}
u_i'&=&\begin{cases}\#m+k+2&\text{ if }u_i=\#m \text{ and }
i+m>k+1,\\
u_i&\text{ otherwise,}\end{cases}
\\
X'&=&X, \text{ except that all jumps in $X$ that pass
$(n+1)\rlh;u_1;...;u_k;\rlt$}\\
&&\phantom{X, } \text{ are raised with $k+2$.}
\end{eqnarray*}
With these two equations all rigid
loops can be removed in $(u_1;...;u_k)^\omega$,
and defining
\[\pgarltpgap(X)=X~\text{ if $X$ is a PGA-program}\]
completes the definition of this pure projection.

We first argue that the expansion equation \eqref{twee} is sound for
the finite case.
Let 
\begin{eqnarray*}
t_1&=&v_1;...;v_r;(n+1)\rlh;u_1;...;u_k;\rlt;w_1;...;w_s,\\
t_2&=&v_1';...;v_r';\#1;u_1';...;u_k';\#1;n\rlh;u_1;...;u_k;\rlt;
w_1;...;w_s.
\end{eqnarray*}
We show that $\pgarltpga(t_1)=_{be}\pgarltpga(t_2)$
by case distinction on the various instructions in $t_1$,
assuming $t_1$ contains $l$ rigid loops with their closure
instructions at positions $i_1,...,i_l$ (so $i_1=r+k+2$).
Without loss of generalization we further assume that
jumps outside the program are such that in
$t_1;\#0;\#0$
they end in one of the latter two $\#0$ instructions, and thus
we can and will leave out the repetition in $\pgarltpga(t_1)$. By a similar
argument, the repetition in $\pgarltpga(t_2)$ is left out.

With respect to the instructions $v_i$, the only interesting case is 
$v_i=\#j$ with $i+j>r$. We distinguish four sub-cases:
\begin{enumerate}
\renewcommand{\theenumi}{\alph{enumi}}
\item
If $i+j=r+1$, this prescribes a jump (via $\#1$) to the instruction
$\psi_{r+2}(u_1)$. In
$t_2$'s projection there is a jump to the instruction
$\psi_{r+2}(u_1')$. We proceed with this case below.
\item
If $r+1<i+j<r+k+2$, then $\psi_q(u_p)$ in $t_1$'s projection and the
associated $\psi_q(u_p')$ in $t_2$'s projection have to be related.
We proceed with this case below.
\item
\label{a}
If $i+j=r+k+2$, then $\pgarltpga(t_1)$ is further determined by
\begin{eqnarray*}
&&\#1;\psi_{r+2}(u_1);...;\psi_{r+k+1}(u_k);\uni{(...)};\\
&&\psi_{r+k+3}(w_1);...;\psi_{r+k+s+2}(w_s);\#0;\#0\use{\rlc:r+k+2} \LC_{r+k+2}(n-1)
...\end{eqnarray*}
and so is $\pgarltpga(t_2)$ (although all its $\psi$-indices and
foci- and counter-indices are
raised with $k+2$, but this is not significant).
So in this case, $\pgarltpga(t_1)=_{be}\pgarltpga(t_2)$.
\item
\label{b}
If $i+j>r+k+2$, then in both $\pgarltpga(t_1)$ and
$\pgarltpga(t_2)$ this prescribes a jump to the $w$-part or to one of the two
added $\#0$'s and
$\LC_{r+k+2}$ respectively
$\LC_{r+2k+4}$ do not play a role, so also in this case
behavioral equivalence holds.
\end{enumerate}
According to the first two cases it remains to be proved that
\begin{align}
&\nonumber\psi_{r+i+1}(u_i);...;\psi_{r+k+1}(u_k);\uni{(...)};\\
&\nonumber\psi_{r+k+3}(w_1);...;\psi_{r+k+s+2}(w_s);\#0;\#0\use{\rlc:r+k+2}
\LC_{r+k+2}(n) ...
\\
\label{vier}
&=_{be}\\
&\nonumber\psi_{r+i+1}(u_i');...;\psi_{r+k+1}(u_k');\#1;\#1;\psi_{r+k+4}(u_1);
...;\psi_{r+2k+3}(u_k);\uni{(...)};\\
&\nonumber\psi_{r+2k+5}(w_1);...;\psi_{r+2k+s+4}(w_s);\#0;\#0
\use{\rlc:r+2k+4} \LC_{r+2k+4}(n-1) ...
\end{align}
for $i=1,...,k$.
We discuss the following cases:
\begin{enumerate}
\setcounter{enumi}{4}
\renewcommand{\theenumi}{\alph{enumi}}
\item
If $u_i=\#j$ and $i+j=k+1$, then in the lhs above the rigid loop
is restarted at its first instruction
with counter value $n-1$, and so happens in the rhs, so the behavioral
equivalence in \eqref{vier} holds.
\item
If $u_i=\#j$ and $i+j>k+1$, this prescribes in both sides a jump to
the $w$-part or to one of the added $\#0$'s,
and the behavioral equivalence in \eqref{vier} holds.
\item
If $u_i=m\rlh $ (and its closure instruction is in the
$u$-part), then in both the lhs and the rhs that rigid loop is
either completed and behavior proceeds while the index $i$ in
\eqref{vier} has raised, or the loop is jumped out and
the resulting position
either matches one of the two cases above, or is into the
$u$-part. In the latter case, 
also the index $i$ in \eqref{vier} has raised.
\end{enumerate}
It follows that for all instantiations of $u_i$
we either obtain the behavioral equivalence in \eqref{vier}, or 
the index $i$ raises until we are at least at position $r+k+2$ 
and behavioral equivalence then follows from the sub-cases \eqref{a} and
\eqref{b} above.
This completes our argument on the soundness of equation \eqref{twee} for
the finite case.
A comparable, but more simple analysis reveals the soundness
of equation \eqref{een} for finite PGArl-programs.

The iterative case is slightly more
complex, as jumps can have a backward target. However, a similar
analysis shows that also in
this case both equations \eqref{een} en \eqref{twee} are sound.
This completes our argument on the soundness of the pure projection
$\pgarltpgap$.

The pure projection clearly provides a combinatorial explosion. 
It can
be concluded that the loop counter service based projection 
$\pgarltpga$ is indeed the best candidate for a defining
semantics: it satisfies both
the criterion \emph{normative semantic adequacy} and
the criterion
\emph{indicative algorithmic adequacy} while the pure projection
satisfies only the first one.

The projection $\pgarltpga$ \emph{defines} the meaning of rigid loop
instructions also for the degenerate case that
a rigid loop header instruction has no associated closure
instruction or vice versa: such a lonely instruction acts as a skip
(i.e., $\#1$).
Finally, note that a rigid loop body of length 0 
is unproblematic: it has no behavioral impact (of course, this 
holds as well for the pure projection).

\section{Conclusions}
\label{Conc}

First we note that the defining projection $\pgarltpga$ uses
finite state services. Indeed, any PGArl-program not containing
repetition can be expanded to one without rigid loops (using the
expansion equations~\eqref{een} and \eqref{twee}).

Although rigid loops are less expressive than arbitrary loops
and fail to
express all finite state threads they can be proven sufficient
for programming state transformations on finite Maurer computers
(see \cite{BM05a,BM05b,BM06}).
Admittedly one may be forced into using quite large loop
counters but in principle it works.

\bigskip

\noindent
\emph{Acknowledgement.}
We thank Bob Diertens for valuable remarks.

\end{document}